\documentstyle[12pt,psfig]{article}
         \makeatletter
         \def\thefigure{\@arabic\c@figure}\def\fps@figure{tbp}
         \def\ftype@figure{1}\def\ext@figure{lof}
         \def\fnum@figure{\protect\footnotesize Fig.\ \thefigure}
         \def\thetable{\@arabic\c@table}
         \def\fps@table{tbp}\def\ftype@table{2}\def\ext@table{lot}
         \def\fnum@table{\protect\footnotesize Table \thetable}
         \def\@listI{\leftmargin\leftmargini\parsep=0pt\itemsep=0pt}
         \def\thebibliography#1{\section{References}\vspace*{-10pt}\list
          {[\arabic{enumi}]}{\settowidth\labelwidth{[#1]}\leftmargin\labelwidth
          \advance\leftmargin\labelsep
          \usecounter{enumi}}
          \def\newblock{\hskip .11em plus .33em minus .07em}
          \sloppy\clubpenalty4000\widowpenalty4000
          \sfcode`\.=1000\relax}
         
         \def\@nomath#1{\ifmmode \fi}
         \def\thesection {\arabic{section}}
         \def\section#1{\addtocounter{section}{1}\setcounter{subsection}{0}
              \bigskip\medskip{\noindent\bf\thesection.\ #1}
              \medskip}
         \def\thesubsection {\arabic{section}.\arabic{subsection}}
         \def\subsection#1{\addtocounter{subsection}{1}
              \medskip{\noindent\thesubsection.\ #1}
              \medskip}
         \makeatother
\begin{document}
\vspace*{0.3in}
\begin{center}
  {\bf The Role of Phase Space in Complex Fragment Emission from Low to
Intermediate Energies}\\
  \bigskip
  \bigskip
  L.G. Moretto, R. Ghetti, K. Jing, L. Phair, K. Tso, G.J. Wozniak\\
  {\em Nuclear Science Division \\
Lawrence Berkeley National Laboratory, Berkeley, CA 94720\\}
  \bigskip
\end{center}
\smallskip
{\footnotesize
\centerline{ABSTRACT}
\begin{quotation}
\vspace{-0.10in}
The experimental emission probabilities of complex fragments by low energy
compound nuclei and their dependence upon energy and $Z$ value are compared to
the transition state rates.
Intermediate-mass-fragment multiplicity distributions for a variety of
reactions at intermediate energies are shown to be binomial and thus reducible 
at all measured
transverse energies. From these distributions a single binary event probability
can be extracted which has a thermal dependence. A strong thermal signature is
also found in the charge distributions. The $n$-fold charge distributions are
reducible to the 1-fold charge distributions through a simple scaling
dictated by fold number and charge conservation.
\end{quotation}}
 
\section{Transition State Rates and Complex Fragment Decay Widths}

The rates for fission decay, as well as for chemical reactions, are calculated
most often by means of the transition state method \cite{Wig38}. In this
approach, the reaction rate is equated to the flux of phase space density
across a ``suitably'' located hyperplane normal to the ``reaction coordinate''.
The ``suitable'' location is typically chosen at a saddle point in collective
coordinate space, which corresponds to a bottleneck in phase space. A smart
choice of the transition state location should minimize the number of phase
space trajectories doubling back across the hyperplane.

The surprising success of the transition state method has prompted attempts to
justify its validity in a more fundamental way, and to identify regimes in
which deviations might be expected \cite{Kra40,Gra83,Han90}. In what follows we
shall compare experimental decay rates for complex fragment emission with
transition state predictions, and search for energy $E$ and atomic number $Z$
dependent deviations that can be expected to exist. 

The transition state expression for the fission decay width is:
\begin{equation}
\Gamma _f=\frac{1}{2\pi\rho (E)}\int\rho
^*(E-B_f-\epsilon)d\epsilon\approx\frac{T_f}{2\pi}\frac{\rho ^*(E-B_f)}{\rho
(E)},
\end{equation}
where $\rho (E)$ is the level density of the compound nucleus, $\rho
^*(E-B_f-\epsilon)$ is the level density at the saddle point, $B_f$ is the
fission barrier, $\epsilon$ is the kinetic energy over the saddle along the
fission coordinate and 
\begin{equation}
\frac{1}{T_f}=\left.\frac{\partial\left[\ln\rho ^*(x)\right]}{\partial
x}\right|_{E-B_f}.
\label{hbar_omega}
\end{equation}

For the one dimensional case in which the only degree of freedom treated
explicitly is the reaction coordinate, the decay width takes the form:
\begin{equation}
\Gamma _f=\hbar\omega\frac{\rho ^*(E-B_f)}{\rho ^*(E)}\approx\hbar\omega
e^{-B_f/T},
\label{emission_rate}
\end{equation}
where $T$ is the temperature of the transition state. Now both level densities
correspond to the same number of degrees of freedom. The quantity $\hbar\omega$
is the oscillator phonon associated with the ground state minimum.

The emission of complex fragments can be treated in an analogous fashion by
introducing the ridge line of conditional saddle points \cite{Moretto75}. Each
mass or charge emission can be associated with a conditional barrier. These
barriers can be measured with techniques similar to those used to determine
fission barriers \cite{McM85}. Recently, nearly complete ridge lines have been
determined for several nuclei: $^{75}$Br \cite{Del91}, $^{90,94}$Mo
\cite{Jing94},
and $^{110,112}$In \cite{McM85}.

The emission rate of a fragment of a given mass or charge can still be
described by an expression similar to that of Eq.~(\ref{emission_rate}). The
quantity $B_f$ becomes the conditional barrier $B_Z$; but what is now the
meaning of $\hbar\omega$? Is there a single value of $\hbar\omega$ for all the
channels or has each channel its own characteristic frequency? We shall
endeavor to answer this question experimentally.

An additional aspect of the problem has been studied by Kramers in his seminal
work \cite{Kra40}. Kramers considered the diffusion of the system from the
reactants' region to the products' region from the point of view of the
Fokker-Planck equation. The new parameter entering the problem is the
viscosity coefficient, which couples the reaction coordinate to the heat bath.
The stationary current solution found by Kramers leads to expressions for the
reaction rates similar to that of the transition state theory, differing only
in the pre-exponential factor, which now includes the viscosity. More recent
work on the same equation has shown that if the system is forced to start at
time $t$=0 at the ground state minimum, a transient time $\tau _f$ exists
during which the reaction rate goes from zero to its stationary value
\cite{Gra83}. Both effects would decrease the overall fission rate compared to
the transition state prediction.

These effects have been advocated as an explanation for the large number of
prescission neutrons observed in the fission of many systems
\cite{Gav81,Hol83,Hin92,Hil92,Hin86}, in apparent contradiction with the
predictions of the transition state method \cite{Hin92,Hil92}. The prescission
neutrons can be emitted either before the system reaches the saddle point, or
during the descent from saddle to scission. Only the former component, however,
has any bearing on possible deviations of the fission rate from its
transition state value, and the separation of the two components is very
difficult indeed.

Furthermore, it has been suggested that the viscosity and the transient time
may depend on the collectivity of the reaction coordinate \cite{Hil89}. More
specifically, the reaction coordinate for a very asymmetric decay should have
little collectivity, while that for a symmetric decay should be very
collective. A study of prescission particles as a function of the size of the
emitted fragment claims to have observed such and effect
\cite{Hin92,Hil89,Mor91}.

We are going to show that the presence or absence of the effects discussed
above can be directly observable in the excitation functions for the emission
of fragments with different $Z$ values. Our procedure uses the transition state
prediction as a null hypotheses, and involves only the replotting of
experimental data without using any specific model. The cross section for the
emission of a fragment of a given $Z$ value can be written as:
\begin{equation}
\sigma _Z=\sigma _0\frac{\Gamma _Z}{\Gamma _T}=\sigma _0\frac{\Gamma _Z}{\Gamma
_n+\Gamma _p+...},
\label{sigmaZ}
\end{equation}
where $\sigma _0$ is the compound nucleus formation cross section and $\Gamma
_T$, $\Gamma _n$, $\Gamma _p$, $\Gamma _Z$ are the total-, 
neutron-, proton-, and $Z$-decay widths, respectively. Notice that $\Gamma _T$
is essentially independent of $Z$
if we confine our observations 
to the excitation energy region where the complex fragment 
emission probability is small.

We now rewrite Eq.~(\ref{sigmaZ}) as follows:
\begin{equation}
\frac{\sigma _Z}{\sigma _0}\Gamma _T\frac{2\pi\rho (E-E_r^{gs})}{T_Z}=\rho
(E-B_Z-E_{r,Z}^s),
\label{rhostar}
\end{equation}
where $T_Z$ is the temperature at the conditional saddle point and 
$E_r^{gs}$, $E_{r,Z}^s$ are the ground state and 
saddle point rotational energies. 
In this way, the left hand side of the equation contains the complex 
fragment cross section which can be measured, 
and other calculable quantities that do not depend 
on $Z$, except $T_Z$ which is only weakly dependent on $Z$. 
The right hand side contains only the level 
density at the conditional saddle calculated 
at the intrinsic excitation energy over the conditional 
saddle, which is calculable if the barrier height is known.

By using the standard Fermi gas level density expression, 
one can rewrite Eq.~(\ref{rhostar}) in the 
following way which takes out the $A$-dependence of the level density:
\begin{equation}
\frac{\ln\left[\frac{\sigma _Z}{\sigma _0}\Gamma _T\frac{2\pi\rho 
(E-E_r^{gs})}{T_Z}\right]}{2\sqrt{a_n}}=\frac{\ln
R_f}{2\sqrt{a_n}}=\sqrt{\frac{a_Z}{a_n}(E-B_Z-E_r^s)},
\label{Rf}
\end{equation}
where $a_Z, a_n$ are the saddle and ground state 
level density parameters. A plot of the left hand side of 
this equation versus the square root of the 
intrinsic excitation energy over the saddle should give a 
straight line, and the slope of this straight 
line should give the square root of $a_Z/a_n$.

\begin{figure}
\centerline{\psfig{file=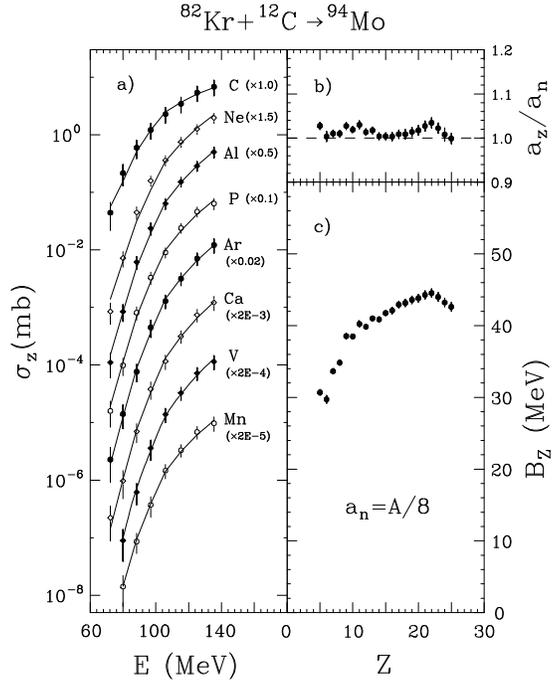,height=9.0cm,angle=180}}
\caption{a) The excitation functions (cross sections vs excitation energy) 
for complex fragments of some typical $Z$ values 
emitted from the compound nucleus $^{94}$Mo produced in the 
reaction $^{82}$Kr+$^{12}$C at beam energies ranging from 6.2 to 
12.2 MeV/u. b) The $a_Z/a_n$ values and c) conditional barriers 
$B_Z$, both extracted by fitting the excitation functions with a 
transition state formalism. The solid lines in a) correspond to 
the fit using an energy level parameter $a_n=A/8$.}
\label{figKr}
\end{figure}

Recently, the excitation functions for a large number of fragment $Z$ 
values have been 
measured for the following systems: $^{75}$Br \cite{Del91}, $^{90,94}$Mo 
\cite{Jing94}, and $^{110,112}$In \cite{McM85}. 
The corresponding conditional barriers have been extracted 
by fitting the excitation functions with the transition state 
formalism. A level density parameter $a_n=A/8$ was assumed in the fitting. 
As an example, Fig.~\ref{figKr}a 
shows some of the excitation functions for fragments with $Z$-values 
from 5 to 25 for the compound 
nucleus $^{94}$Mo. 
The solid lines in Fig.~\ref{figKr}a correspond 
to the best fit to the experimental data. The 
extracted ratios $a_Z/a_n$ are close to unity for all $Z$ values 
(see Fig.~\ref{figKr}b). The 
extracted conditional 
barriers increase from 30-45 MeV as the 
charge of the emitted fragment increases (see Fig.~\ref{figKr}c).

\begin{figure}
\centerline{\psfig{file=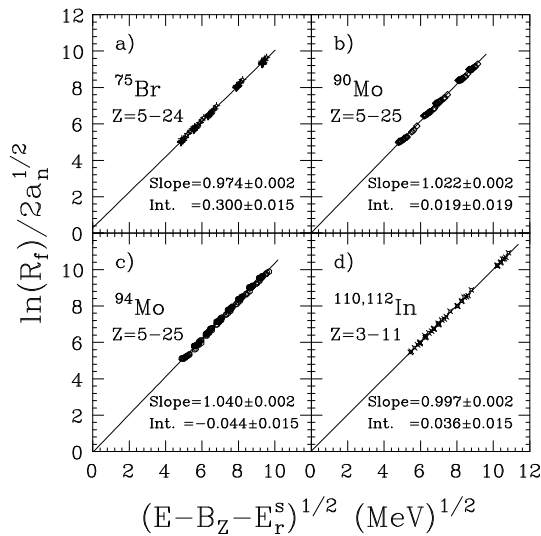,height=7.0cm,angle=180}}
\caption{$\ln R_f/2\protect\sqrt{a_n}$ (see Eq.~(\protect\ref{Rf}) 
and text) vs the square root of 
the intrinsic excitation energy for four compound nuclei: $^{75}$Br a), 
$^{90}$Mo b), $^{94}$Mo c), and $^{110,112}$In d).  All the excitation 
functions for the indicated $Z$ range are included for each 
compound nucleus. The solid lines are the linear fits to the 
data. The error bars are smaller than the size of symbols.}
\label{reduce_cf}
\end{figure}

Eq.~(\ref{Rf}) suggests that it should be possible 
to reduce {\em all} the excitation functions for the 
emission of different complex fragments from a 
given system to a single straight line. In Fig.~\ref{reduce_cf} all 
the excitation functions associated with each of four compound nuclei 
($^{75}$Br, $^{90}$Mo, $^{94}$Mo, and 
$^{110,112}$In) are plotted according to Eq.~(\ref{Rf}). 
There are 20, 21, 21, and 9 excitation functions for $^{75}$Br, 
$^{90}$Mo, $^{94}$Mo, and $^{110,112}$In, respectively. 
We see that all the excitation 
functions for each $Z$-value 
fall with remarkable precision on the same line 
that is in fact straight, has a slope near unity and 
passes closely through zero.  

%

\begin{figure}
\centerline{\psfig{file=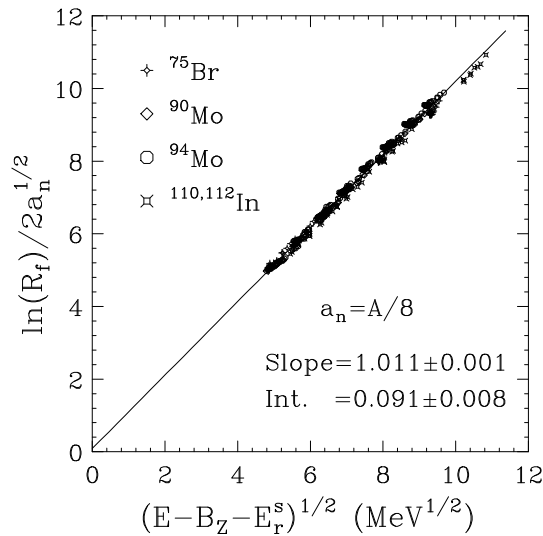,height=7.0cm,angle=180}}
\caption{The same as Fig.~\protect\ref{reduce_cf} with data for 
all four nuclei in a single plot. 
The straight line is the linear fit to all the data points.}
\label{collapse_all}
\end{figure}

As a final virtuoso touch, we can try to collapse ALL 
the excitation functions for ALL $Z$ 
values and for all compound nuclei into a single straight line. 
The resulting plot for four different 
compound nuclei is shown in Fig.~\ref{collapse_all}. 
It includes a total of 71 excitation functions spanning a $Z$ 
range from 3 to 25. The collapse of all 
the experimental excitation functions for all of the different 
$Z$-values and all the systems onto a 
single straight line is strong evidence for the validity of the 
transition state formalism and for the absence of $Z$- and 
$E$-dependent deviations. In particular, one 
is led to the following conclusions:
a) Once one removes the phase space 
associated with the non reactive degrees of freedom at the 
conditional saddle point, the reduced rates are IDENTICAL for fragments of 
all $Z$-values. 
Within the experimental sensitivity, the quantity $\hbar\omega$
in Eq.~(\ref{hbar_omega}) appears to be $Z$ independent.
b) For all fragments, there is no 
deviation from the expected linear dependence over the 
excitation energy range from 50-130 MeV explored. 
This seems to rule out, for 
all $Z$-values, 
transient time effects which should 
become noticeable with increasing excitation energy.
c) The slope, which corresponds to the $\sqrt{a_Z/a_n}$, 
is essentially 1 for all $Z$ values of all systems 
studied. 
d) The intercept of the straight line, 
which is associated with the channel frequency $\omega$, is 
essentially zero and shows no obvious dependence on the fragment $Z$-values 
(i.e., the 
collectivity).

We conclude that in this extended data set 
there is no evidence for transient effects either 
directly or through their expected 
dependence upon the mass of the emitted fragment. Furthermore 
it appears that the channel frequency 
is the same for all the different $Z$ decay channels.

\section{Reducibility at intermediate energies}

At low excitation energies, complex fragments are emitted with low probability
by a compound nucleus mechanism \cite {Sob83,Moretto88}. At increasingly 
larger energies, the
probability of complex fragment emission increases dramatically, until several
fragments are observed within a single event \cite 
{Gue89,Gro90,Moretto93}. The nature of this
multifragmentation process is at the center of much current attention. For
example, the timescale of fragment emission and the associated issue of
sequentiality versus simultaneity are the objects of intense theoretical
and experimental 
study. 

Recent experimental work \cite {Delis93,Pou93} has shown that 
the excitation functions for
the production of two, three, four, etc. fragments give a characteristically
linear Arrhenius plot, suggesting a statistical 
energy dependence. 

A fundamental issue, connected in part to those mentioned above, is that of
reducibility: can multifragmentation be reduced to a combination of (nearly)
independent emissions of fragments? More to the point, can the probability for
the emission of $n$ fragments be reduced to the emission probability of just one
fragment?

Recently, it has been experimentally observed in many 
reactions that for
any value of the transverse energy $E_t$, the $n$-fragment emission 
probability $P_n$ is 
reducible to the one-fragment emission probability $p$ through a binomial
distribution \cite {Moretto95,Tso95}

\begin{equation}
P_{n}^{m}=\frac{m!}{n!(m-n)!}p^n(1-p)^{m-n}.
\label{binomial}
\end{equation}

\begin{figure}
\vspace*{5.0in}
\includegraphics{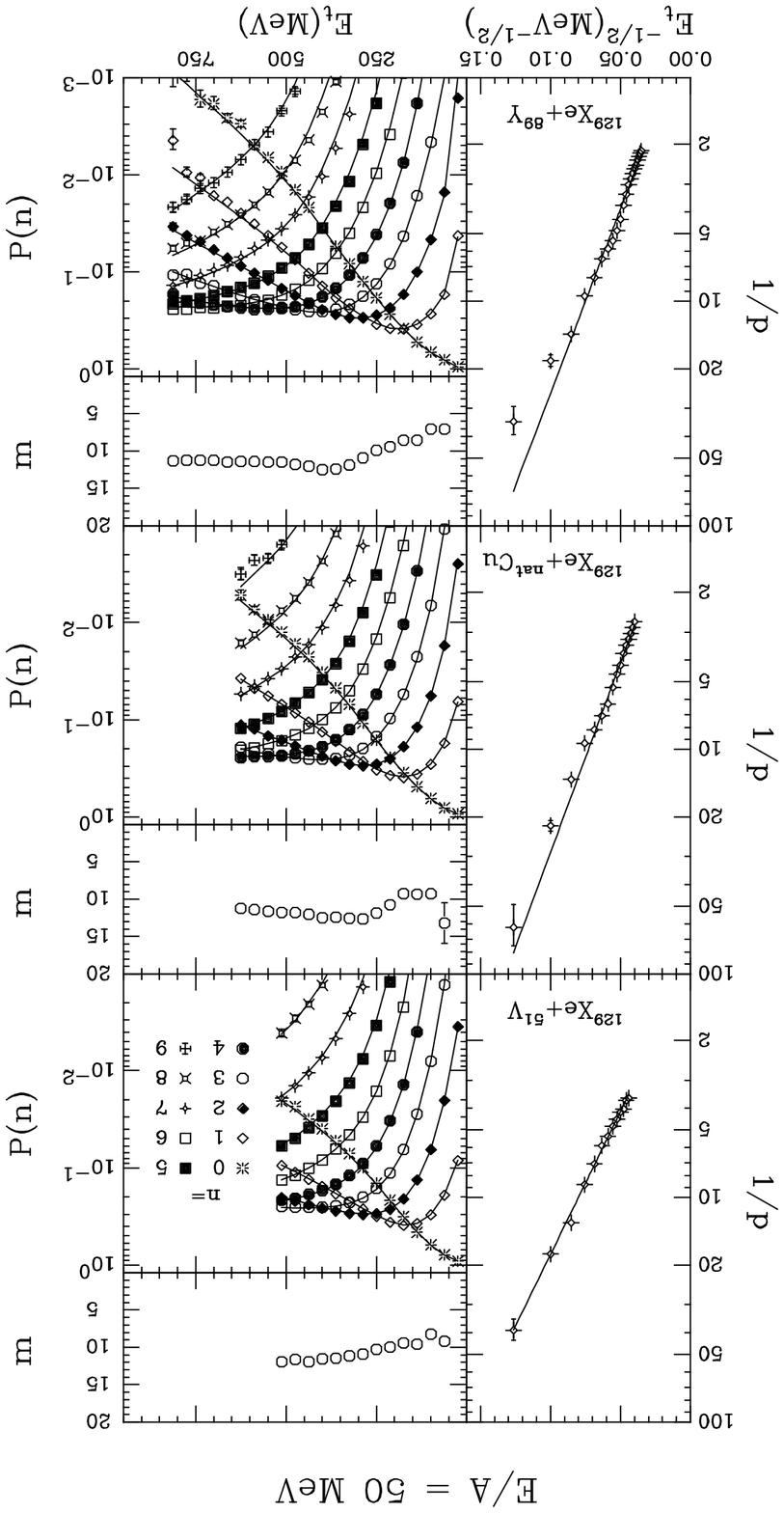} 
\includegraphics{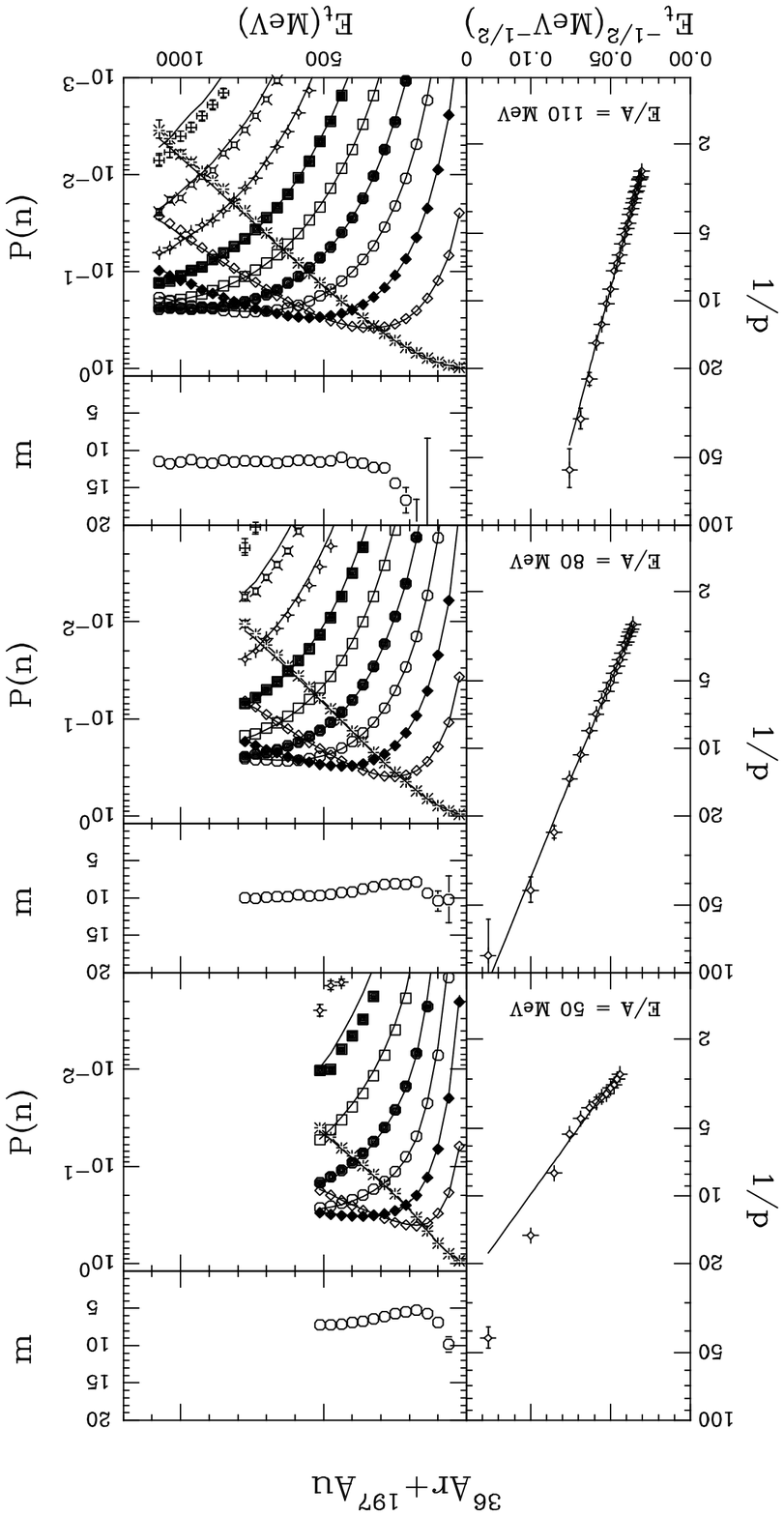} 
    \caption[]{\footnotesize
For the $^{129}$Xe induced reactions at $E/A$=50 MeV (left figure) and the 
$^{36}$Ar+$^{197}$Au reactions at $E/A$=50, 80 and 110 MeV (right figure): 
(left panel) the
reciprocal of the single fragment emission probability $1/p$ as a function of
$1/\sqrt{E_t}$; and (right panel) the parameter $m$ (number of the throws in
the binomial distribution) and the probability $P(n)$ of emitting $n$
intermediate mass fragments (IMF, $3\le Z\le 20$) as a function of the
transverse energy $E_t$. The solid lines through the excitation functions
correspond to binomial distributions calculated with the given values of $m$
and $p$. (See Eq. (\protect\ref{binomial}))}
\label{binomial_red}
\end{figure}

This empirical evidence indicates that multifragmentation can be thought of as
a special combination of nearly independent fragment emissions. The binomial
combination of the elementary probabilities points to a combinatorial structure
associated with a time-like or space-like one-dimensional sequence. It was also
found that the log of such one-fragment emission probabilities ($\log p$) 
plotted vs $1/\sqrt{E_t}$ (Arrhenius plot) gives a 
remarkably straight line. This linear dependence is
strongly suggestive of a thermal nature for $p$,

\begin{equation}
p=e^{-B/T}
\label{littlep}
\end{equation}
under the assumption that the temperature $T\propto \sqrt{E^{*}}$ where $E^{*}$ 
is the excitation energy. Examples of the binomial decomposition of the
$n$-fragment emission probabilities $P_n$ into a one-fragment emission
probability $p$, and the resulting Arrhenius plot for $p$ is given in Fig.
\ref{binomial_red}. The extraordinary
quantitative agreement between the calculations and the experimental data
confirms the binomiality of the multifragmentation process.

The more directly interpretable physical parameter contained in this analysis
is the binary barrier $B$ (proportional to the slope of the Arrhenius plot in 
Fig. \ref{binomial_red}). One
may wonder why a single binary barrier suffices, since mass asymmetries with
many different barriers may be present. Let us consider
a barrier distribution as a function of mass asymmetry $x$ of the form 
$B=B_0+ax^n$, where $B_0$
is the lowest barrier in the range considered. Then,
\begin{equation}
p=\frac{\Gamma}{\hbar\omega_0}=\int{e^{-B_0/T}e^{-ax^n/T}dx}\cong
\left( \frac{T}{a}  \right)^{1/n}e^{-B_0/T}
\end{equation}
Thus the simple form of Eq.~(\ref{littlep}) 
is retained with a small and renormalizable
pre-exponential modification.

One possible interpretation of the reducibility discussed above is sequential
decay with constant probability $p$. Assuming that the (small) fragments, once
produced do not generate additional fragments or disappear, the binomial
distribution follows directly. In this framework, it is possible to translate
the probability $p$ into the mean time separation between fragments. In other
words, we can relate the $n$-fragment emission probabilities to the mean time
separations between fragments. The validity of this interpretation is  testable
by experiment.

We have tried to find alternative explanations to the sequential description
for the binomial distributions with thermal probabilities. An obvious model is
a chain of $m$ links with probability $p$ that any of the links is broken. The
probability that $n$ links are broken is given by Eq. (\ref{binomial}). 
This result is, of
course, strictly dependent on the dimensionality of the model, and its
relevance to multifragmentation is unclear. Nevertheless, it stresses again the
{\em fundamental reducibility} of the multifragmentation probability to a binary
breakup probability $p$.


\section{Charge Distributions}

These aspects of reducibility and thermal scaling in the integrated fragment
emission probabilities lead naturally to the question: Is the charge
distribution itself reducible and scalable? In particular, what is the charge
distribution form that satisfies the condition of reducibility and of thermal
dependence?

Let us first consider the aspect of reducibility as it applies to the charge
distributions. In its broadest form, reducibility demands that the probability 
$p(Z)$, from which an event of $n$ fragments is generated by $m$ trials, is 
the same at 
every step of extraction. The consequence of this extreme reducibility is
straightforward: the charge distribution for the one-fold events is the same as
that for the $n$-fold events and equal to the singles distributions, i.e.:

\begin{equation}
P_{(1)}(Z)=P_{(n)}(Z)=P_{singles}(Z)=p(Z).
\label{reducible}
\end{equation}

We now consider the consequences of the thermal dependence of $p$ 
\cite {Moretto95} on the charge
distributions. If the one-fold = $n$-fold = singles distributions 
is thermal, then

\begin{equation}
P(Z)\propto e^{-\frac{B(Z)}{T}}
\end{equation}
or $T\ln P(Z)\propto-B(Z)$. This suggests that, under the usual 
assumption $E_t\propto E^*$ \cite {Moretto95}, the function
\begin{equation}
\sqrt {E_t}\ln {P(Z)}=D(Z)
\end{equation}
should be independent of $E_t$.

\begin{figure}
\centerline{\psfig{file=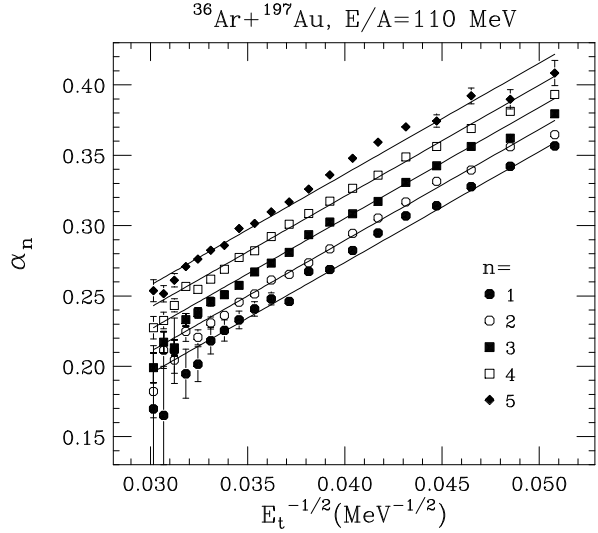,height=7.0cm,angle=90}}
    \caption[]{\footnotesize
	The exponential fit parameter $\alpha _n$ 
(from fits to the charge distributions,
see Eq.(\protect \ref{fitP(Z)})) is plotted as a function of 
1/$\protect \sqrt{E_t}$. 
The solid lines are a fit to the values of $\alpha _n$ using Eq.
(\protect \ref{alpha_fit}). }
\label{alpha_n}
\end{figure}

In the $^{36}$Ar+$^{197}$Au reaction, as in other 
reactions \cite {Kim92,Bow92}, the IMF
charge distributions are empirically found to be nearly exponential functions
of $Z$
\begin{equation}
P_n(Z)\propto e^{-\alpha _nZ}.
\label{fitP(Z)}
\end{equation}
In light of the above considerations, we would expect for 
$\alpha _n$ 
the following simple dependence
\begin{equation}
\alpha _n\propto \frac{1}{T} \propto \frac{1}{\sqrt{E_t}}
\label{thermal}
\end{equation}
for all folds $n$. Thus a plot of $\alpha _n$ vs $1/\sqrt{E_t}$ should 
give nearly straight lines. This is
shown in Fig. \ref{alpha_n} for $^{36}$Ar+$^{197}$Au at $E/A$=110 MeV.

The expectation of thermal scaling appears to be met quite satisfactorily. For
each value of $n$ the exponent $\alpha _n$ shows the linear dependence on 
$1/\sqrt{E_t}$ anticipated in Eq. (\ref{thermal}). 
On the other hand, the extreme reducibility condition demanded by Eq. 
(\ref{reducible}),
namely that $\alpha _1=\alpha _2=...=\alpha _n=\alpha$, is not rigorously 
met. Rather 
than collapsing on a single straight line, the
values of $\alpha _n$ for the different fragment multiplicities are 
offset one with
respect to another by what appears to be a small constant quantity.

In fact, one can fit all of the data remarkably well, assuming for $\alpha _n$ 
the form:
\begin{equation}
\alpha _n=\frac{K'}{\sqrt{E_t}}+nc
\label{alpha_fit}
\end{equation}
which implies:
\begin{equation}
\alpha _n=\frac{K}{T}+nc
\end{equation}
or more generally, for the $Z$ distribution:
\begin{equation}
P_n(Z)\propto e^{-\frac{B(Z)}{T}-ncZ}.
\label{Pn}
\end{equation}
Thus, we expect a more general reducibility expression for the charge
distribution of any form to be:
\begin{equation}
\left[ \ln P_n(Z)+ncZ \right] \sqrt {E_t}=F(Z)
\label{reduced}
\end{equation}
for all values of $n$ and $E_t$. 
This equation indicates that it should be possible 
to reduce the 
charge distributions associated with any intermediate mass fragment
multiplicity to the charge distribution of the singles.

What is the origin of the regular offset that separates the curves in
Fig.~\ref{alpha_n}?
The general form of Eq. (\ref{Pn}) suggests the presence of an entropy 
term that does
not depend explicitly on temperature. The general expression for the free 
energy (in terms of enthalpy $H$, temperature $T$ and entropy $S$)
\begin{equation}
\Delta G=\Delta H(Z)-T\Delta S(Z)
\end{equation}
leads to the distribution
\begin{equation}
P(Z)\propto e^{-\frac{\Delta H(Z)}{T}+\Delta S}.
\end{equation}

Typically, $\Delta S$ is of topological or combinatorial origin. For instance,
a factor of this sort would appear in the isomerization of a molecule involving
a change of symmetry.
In our specific case $\Delta S$
 may point to an asymptotic combinatorial structure of the multifragmentation
process in the high temperature limit. As an example, we consider the Euler
problem of an integer to be written as the sum of smaller integers, and
calculate 
the resulting integer distribution. Specifically, let us consider an integer 
$Z_0$ to be broken into $n$ pieces. Let $n_Z$ be the number of pieces 
of size $Z$. It can be shown \cite{Phair95} that
\begin{equation}
n_Z=\frac{n^2}{Z_0}e^{-\frac{nZ}{Z_0}}=cn^2e^{-cnZ}.
\label{Euler}
\end{equation}
This expression has the correct asymptotic structure for 
$T\rightarrow \infty$ required by Eq. (\ref{Pn}).
The significance of this form is transparent: First, the overall scale for the
fragment size is set by the total charge $Z_0$. 
Second, for a specific multiplicity
$n$, the scale is reduced by a factor $n$ to the value $Z_0/n$.

%
%
%

\section{Phase Coexistence}

While Eq. (\ref{Euler}) obviously implies charge conservation, 
it is not necessary that charge conservation be 
implemented as suggested by it. 
In fact it is easy to envisage a regime where the 
quantity $c$ should be zero. 
Sequential thermal emission is a case in point. 
Since any fragment does not know how many 
other fragments will follow its emission, 
its charge distribution can not reflect 
the requirement of an unbiased partition of the total charge among $n$
fragments. We have in mind a liquid drop evaporating fragments of 
different size and binding energy.
``Charge conservation'' will affect the distribution minimally, unless
evaporation consumes the entire system, and even then, not in the sense of an
unbiased partition. 

On the other hand, in a simultaneous emission 
controlled by a $n$-fragment transition state \cite{Lop89,Lop90}, 
fragments would be strongly aware of each other, and 
would reflect such an awareness through the charge distribution.

The question then arises whether $c=0$ or $c>0$, 
or even better, whether one can identify a 
transition from a regime for which $c=0$ to a new regime for 
which $c>0$. In order to answer this question, 
we have studied the charge distributions 
as a function of fragment multiplicity $n$ and transverse energy $E_t$ 
for a number of systems and excitation energies. 
Specifically, we will present data for the reaction $^{36}$Ar+$^{197}$Au at 
$E/A$=80 and 110 MeV and the reaction $^{129}$Xe+$^{197}$Au at 
$E/A$=50 and 60 MeV.


\begin{figure}
\centerline{\psfig{file=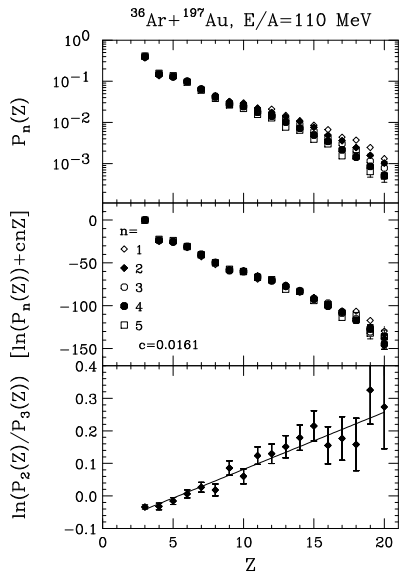,height=7.5cm,angle=180}}
    \caption[]{\footnotesize
Top panel: the $n$ gated charge distributions
$P_n(Z)$ for the reaction $^{36}$Ar+$^{197}$Au at
$E/A$=110 MeV. The charge distributions were constructed from events with
$E_t$=650$\pm$20 MeV and $n$=1-5. 
Middle panel: the ``reduced'' charge distribution
\protect\cite{Phair95} for the same data using the indicated value of $c$. (The
data here are normalized at $Z$=3). Bottom panel: the log of the ratio of 
$P_2(Z)/P_3(Z)$. The slope corresponds to $c$ for $n$=2 
(see Eq.(\protect\ref{c_n})). The statistical error bars are shown for errors
larger than the symbol size.}
\label{reduced_charge}    
\end{figure} 

\begin{figure}
\centerline{\psfig{file=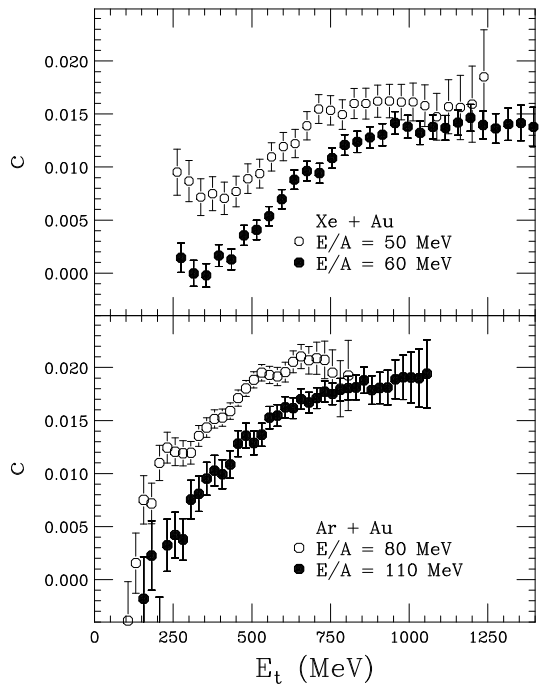,height=9cm,angle=180}}
\caption{Plots of the coefficient $c$ versus $E_t$ for the 
reactions $^{129}$Xe+$^{197}$Au at $E/A$=50 and 60 MeV (top panel) and
$^{36}$Ar+$^{197}$Au at $E/A$=80 and 110 MeV (bottom panel). The error bars are
statistical.
}
\label{chi2_c}
\end{figure}

A general approach for measuring $c$, which does not 
depend on any specific form for the charge distribution, 
is to
construct at each $E_t$ the ratio 
\begin{equation}
\frac{P_n(Z)}{P_{n+1}(Z)}=e^{cZ}. 
\label{c_n}
\end{equation}
A value of $c$
can be extracted for each $n$ 
by taking the log of this ratio and finding the slope of the
resulting graph  
(see bottom panel of Fig. \ref{reduced_charge}). A weighted average (over all
IMF multiplicities $n$) for $c$ can then be constructed at all $E_t$. 
Alternatively,
a $\chi ^2$ can be constructed
in terms of the differences in $F(Z)$ (see Eq. (\ref{reduced})) 
between any pairs of $n$ values 
and minimized as a function of $c$. 
These procedures yield essentially the same results. 
These results are reported in Fig. \ref{chi2_c} for 
the $^{129}$Xe+$^{197}$Au and $^{36}$Ar+$^{197}$Au reactions. 

It is interesting to notice that for all 
reactions and bombarding energies the quantity $c$ starts at or 
near zero,  
it increases with increasing $E_t$ for small $E_t$ values, and seems to 
saturate to a constant value at large $E_t$.

This behavior can be compared to that 
of a fluid crossing from the region 
of liquid-vapor coexistence to the region 
of overheated and unsaturated vapor. 
In the coexistence region, 
the properties of the saturated vapor cannot 
depend on the total mass of fluid. 
The presence of the liquid phase guarantees 
mass conservation at all average densities for 
any given temperature. A change in mean 
density (volume) merely changes the relative amount 
of the liquid and vapor, without altering 
the properties of the saturated vapor. 
Hence the vapor properties, and, in particular, the 
cluster size distributions cannot 
reflect the total mass or even the 
mean density of the system. In our notation, $c=0$.

On the other hand, in the region of unsaturated vapor, 
there is no liquid to insure mass conservation. 
Thus the vapor itself must take care of this conservation, 
at least grand canonically. In our notation, $c>0$.

This description should not be taken too literally, 
for a variety of reasons, one of which is the 
finiteness of the system. 
The $c=0$ regime 
may signify an evaporative-like emission 
from a source which survives as a charge 
conserving residue (liquid), while the $c>0$ 
regime may signify the complete vaporization of the source.

In order to test these ideas for a finite system, 
percolation calculations \cite {Bau88} 
were performed for systems of $Z_0$=97, 160 and 
400 as a function of the percentage of bonds broken ($p_b$) 
in the simulation.
Values of $c$ were extracted (using Eq.(\ref{c_n})) as a function of $p_b$.

\begin{figure}
\centerline{\psfig{file=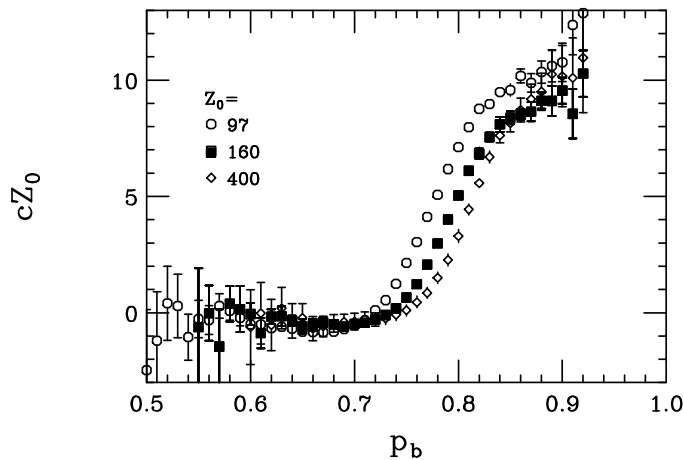,height=6.0cm,angle=90}}
\caption{A plot of $cZ_0$ 
versus the percentage of broken bonds $p_b$ 
from a percolation calculation \protect\cite{Bau88} for three 
systems $Z_0$=97 (circles), $Z_0$=160 (squares) and $Z_0$=400 (diamonds). 
The statistical 
error bars are shown for errors
larger than the symbol size.
}
\label{percolation}
\end{figure}

The results are shown in Fig. \ref{percolation}. Guided by the insight
gleaned from the approximate solution to Euler's problem (see Eq.(\ref{Euler})) 
we have scaled the extracted values of $c$ by the source size $Z_0$ in order to
remove this leading dependence and to evidentiate the true finite size effects.
For values of $p_b$ smaller than the critical (percolating) value 
($p_b^{crit}\approx $ 0.75 for an infinite system), 
we find $c=0$. 
This is the region in which a large (percolating) 
cluster is present. As $p_b$ goes above its critical 
value, the value of $c$ increases, and eventually 
saturates in a way very similar to 
that observed experimentally. Due to the 
finiteness of the system the transition is 
smooth rather than sharp
and can be made sharper by increasing the size of the system. 

Before proceeding, let us remind ourselves that charge conservation is 
{\it not} a
finite-size effect. For instance, the chemical potential, introduced in
statistical mechanics to conserve mass, survives the thermodynamical limit and 
retains its meaning for an infinite
system, despite the fact that the extensive thermodynamic quantities go to
infinity. In our case, while it is true that $c$ goes to zero or that $1/c$
goes to infinity, it is also true that the product $cZ_0$ tends to a finite
limit nearly independent of $Z_0$. 

The significance of the actual experimental value 
of $c$ in the region where it seems to saturate is unclear. 
In Eq.(\ref{Euler}), $c$ takes a direct meaning for the Euler problem: 
$c=1/Z_0$.
It should be noted that our analysis is not directly comparable to the Euler
solution (Eq.(\ref{Euler})) since we have restricted 
ourselves to a limited region
($3\leq Z\leq 20$)
of the total charge distribution for our study of how the source is partitioned
into different IMF multiplicities. 
It must also be appreciated that Eq. (\ref{Euler}) 
and the associated dependence of $c$ upon $Z_0$ 
are characteristic of a one-dimensional percolation model. 
In light of the points mentioned above, it is not unexpected that 
$c$ appears to be proportional, but not equal, to $1/Z_0$ 
in the three-dimensional percolation calculation 
reported in Fig. \ref{percolation}. 
An interpretation of $c$ 
in terms of the source size may be 
possible when more data and a better 
understanding of the percolation of finite systems are available.

%
%
%
%
%
\vspace{0.3 in}
{\bf Acknowledgements}

This work was supported by the Director, Office of Energy Research, Office of
High Energy and Nuclear Physics, Nuclear Physics Division of the US Department
of Energy, under contract DE-AC03-76SF00098 and by the National Science
Foundation under Grant Nos. PHY-8913815, PHY-90117077, and PHY-9214992.

%
%
%
%
%
%
%
%
%
%
%


\begin{thebibliography}{99}
\bibitem{Wig38} E. Wigner, Trans. Faraday Soc. {\bf 34}, 29 (1938).
\bibitem{Kra40} H.A. Kramers, Physica {\bf 7}, 284 {1940}.
\bibitem{Gra83} P. Grange {\em et al.}, Phys. Rev. C {\bf 27}, 2063 (1983).
\bibitem{Han90} P. H\" anggi {\em et al.}, Rev. Mod. Phys. {\bf 62}, 251 (1990).
\bibitem{Moretto75} L.G. Moretto, Nucl. Phys. A {\bf 247}, 211 (1975).
\bibitem{McM85} M.A. McMahan {\em et al.}, Phys. Rev. Lett. {\bf 54}, 1995 
(1985).
\bibitem{Del91} D.N. Delis {\em et al.}, Nucl. Phys. A {\bf 534}, 403 
(1991).
\bibitem{Jing94} L.G. Moretto {\em et al.}, Phys. Rev. Lett. {\bf 74}, 3557
(1995).
\bibitem{Gav81} A. Gavron {\em et al.}, Phys. Rev. Lett. {\bf 47}, 1255 
(1981).
\bibitem{Hol83} E. Holub {\em et al.}, Phys. Rev. C {\bf 28}, 252
(1983).
\bibitem{Hin92} D.J. Hinde {\em et al.}, Phys. Rev. C {\bf 45}, 1229 
(1992).
\bibitem{Hil92} D. Hilscher and H. Rossner, Ann. Phys. Fr. {\bf 17}, 471 
(1992).
\bibitem{Hin86} D.J. Hinde {\em et al.}, Nucl. Phys. A {\bf 452}, 550 
(1986).
\bibitem{Hil89} D. Hilscher {\em et al.}, Phys. Rev. Lett {\bf 62}, 1099 
(1989).
\bibitem{Mor91} E. Mordhorst {\em et al.}, Phys. Rev. C {\bf 43}, 1991 
(1991).
\bibitem{Sob83}  L.G. Sobotka, et. al., Phys. Rev. Lett. {\bf 51}, 2187 (1983).
\bibitem{Moretto88}  L.G. Moretto and G.J. Wozniak, Prog. Part. \& Nucl. 
Phys. {\bf 21}, 401 (1988).
\bibitem {Gue89} D. Guerreau, {\it Formation and Decay of Hot Nuclei: 
The Experimental Situation} ed. (Plenum Publishing Corp., 1989).
\bibitem {Gro90} D.H.E. Gross, Rep. Prog. Phys. {\bf 53}, 605 (1990).
\bibitem {Moretto93} L.G. Moretto and G.J. Wozniak, Ann. Rev. Part. Nucl. 
Sci. {\bf 43}, 379 (1993).
\bibitem {Delis93} L.G. Moretto, D.N. Delis, and G.J. Wozniak, 
Phys. Rev. Lett. {\bf 71}, 3935 (1993).
\bibitem {Pou93} J. Pouliot, et al., Phys. Rev. C {\bf 48}, 2514 (1993).
\bibitem {Moretto95} L.G. Moretto, et al., Phys. Rev. Lett. {\bf 74}, 
1530 (1995).
\bibitem{Tso95} K. Tso et al., Phys. Lett. B {\bf 361}, 25 (1995).
\bibitem{Kim92} Y.D. Kim et al., Phys. Rev. {\bf C45}, 338 (1992).
\bibitem{Bow92} D.R. Bowman et al., Phys. Rev. {\bf C46}, 1834 (1992).
\bibitem{Phair95} L. Phair et al., Phys. Rev. Lett. {\bf 75}, 213 (1995).
\bibitem{Lop89} J.A. Lopez and J. Randrup, Nucl. Phys. A {\bf 503}, 
183-222 (1989).
\bibitem{Lop90} J.A. Lopez and J. Randrup, Nucl. Phys. A {\bf 512}, 
345 (1990).
\bibitem{Bau88} W. Bauer, Phys. Rev. C {\bf 38}, 
1297 (1988).

\end{thebibliography}
\end{document}